\documentclass[letter]{IEEEtran}
\usepackage{array}
\usepackage{float}
\usepackage{textcomp}
\usepackage{amsmath}
\usepackage{graphicx}
\usepackage{algorithm}
\usepackage{algpseudocode}


\begin{document}

\title{Robust and Accurate Global Motion Estimation Using the Student-t Distribution}

\author{Yifan~Zhou$^{\dagger}$~and~Simon~Maskell$^{\dagger}$ \thanks{$^{\dagger}$Yifan~Zhou and Simon~Maskell are with the School of Electrical Engineering, Electronics and Computer Science, University
of Liverpool, UK, e-mail: \protect{\{yifanz, smaskell\}@liverpool.ac.uk}.}}

\maketitle


\markboth{}{}

\begin{abstract}
Pixel-based Global Motion Estimation (GME) has always struggled to simultaneously reject outliers, avoid local minima and run quickly. There are many robust cost functions that perform well in terms of rejecting outliers, but they can yield unstable results during long image sequences as a result of their inability to adjust to changes in image content. In this letter, we propose a parameterised student-t cost function that can interpolate between two cost functions that are amongst the most widely in image registration problems, the L2 norm and the Cauchy-Lorentzian function. We also propose a parameter estimation method that helps to find the best parameters for the proposed cost function. Experiments prove that the proposed approach can estimate global motion accurately relative to the existing cost functions without demanding a higher computational cost.
\end{abstract}





\IEEEpeerreviewmaketitle{}

\section{Introduction}

Global motion estimation aims to find the background motion between two images. It is widely used in video processing to infer camera movement (e.g.,~\cite{hsu2004mosaics}\cite{vella2002digital}). The problem motivating this letter is how best to extract background motion from a video that includes a significant number of pixels that are associated with moving foreground objects and relatively few that are introduced by errors associated with the recording device (e.g., salt and pepper noise and blurring caused by the camera being out of focus). The foreground pixels associated with moving objects are usually significant outliers and are near one another such that it is challenging to eliminate their influence on the estimated motion using standard image processing techniques (e.g., medium or Gaussian filters)~\cite{wei1999pre}. The clusters of pixels associated with each foreground object also have consistent motions relative to the background. This letter aims to solve the problem of GME in the context of such foreground pixels.

Global motion estimation (sometimes called image registration) is not a new topic, and related research is extensive. In general, GME can be categorised into two kinds of approach:  feature-based~\cite{botterill2010real}\cite{nister2004visual} and pixel-based (‘direct’)~\cite{zitova2003image}\cite{szeliski2006image} approaches. The feature-based approaches match features based on the descriptor of features (e.g., SIFT~\cite{lowe1999object}, SURF~\cite{bay2006surf}) and then estimate the global motion given the matched features. Pixel-based approaches use the raw pixel values to estimate global motion while considering all pixels in the image. Several comparative studies have been conducted in the context of these approaches (see~\cite{szeliski2006image} or \cite{haller2009evaluation}). These studies conclude that the pixel-based approach is more accurate and robust to poor frame quality. We will focus on the pixel-based approach in this letter.

Modern pixel-based approaches are based on the method developed by Lucas and Kanade~\cite{lucas1981iterative}. There are numerous extensions that have been proposed over recent decades. For example,~\cite{baker2004lucas}~is a well-cited review paper that describes many of these extensions and lists their advantages. In the context of the original Lucas-Kanade approach, the L2 norm (or the quadratic cost function) is used to measure the error between two images. The algorithm exploits the fact that the L2 norm is both continuous and has a smooth first derivative~\cite{lucas1981iterative}. It offers some robustness to small quantities of outliers and noise. However, several papers have identified that pronounced noise and outliers can degrade the estimation when using the L2 norm. There are  several papers that propose replacing the L2 norm with other robust cost functions, as commonly used in the statistical community. Examples include the L1 norm, Huber's M-estimator, Tukey's M-estimator, and the Cauchy-Lorentzian function. These cost functions all use ‘soft’ gates in the optimisation process (e.g.,~\cite{stewart1999robust, black1996robust, yap2009nonlinear, arya2007image, black1993framework}). Indeed, these cost functions offer improved robustness to the presence of outliers by constraining or reducing the contribution to the cost function made by pixels associated with significant errors. However, they all have some disadvantages regarding either accuracy or computational efficiency. For example, using the L1 norm can be considered to be related to using the median estimator, which is much less sensitive to errors than a mean estimator (which is related to using the L2 norm). However, due to the discontinuous first differential of the L1 norm, minimising the L1 norm is more challenging than minimising the L2 norm. Although there are several ways to optimise the error function (e.g., \cite{yap2009nonlinear}), using an L1 norm still suffers from slower convergence and struggles to respond well to poor initialisation. Huber's M-estimator limits the contribution to the cost function made by pixels with high errors, but the high-error pixels still have higher influence on Huber's M-estimator than on the other robust cost functions listed we discuss herein. This limits the extent to which Huber's M-estimator is robust to the presence of outliers. Tukey's M-estimator adopts a more aggressive approach to limiting the contribution made by high-error pixels. However, some of the pixels that are currently considered to have high errors could be inliers with respect to the final estimate. This makes the optimisation process prone to falling into local minima and can reduce computational efficiency. The Cauchy-Lorentzian function offers good performance relative to the aforementioned functions. The corresponding cost functions are shown in Table~\ref{tab:curves} together with the corresponding influence functions (as discussed in more detail in Section~II).

The student t-distribution is a widely used robust function in the context of statistics (e.g., see in~\cite{marrs2002expected}). In the context of image registration, previous work has capitalised on the heavy-tailed and parameterised nature of the distribution. Previous work has considered a complex optimisation of a mixture of student t-distributions (e.g.,~\cite{gerogiannis2009mixtures}\cite{zhou2014robust}). However, these methods use EM to build the mixture that considers all the pixels and are therefore sufficiently slow that application to video is problematic. In this letter, we will focus on processing videos. More specifically, we propose to adapt the parameters of a student-t in response to a sequence of images such that the GME method provides a cost function that interpolates between the L2 norm and the Cauchy-Lorentzian function.

This letter is organised as follows: Section~II describes the proposed parameterised student-t function. Section~III describes how to estimate the affine transformation using the proposed cost function. The experiments are reported in Section~IV and Section~V concludes this letter.

\begin{table}[htp]
\caption{Curves Of Existing Cost Functions \label{tab:curves}}
\begin{centering}
\begin{tabular}{|>{\centering}m{1.5cm}|>{\centering}m{2.8cm}|>{\centering}m{2.8cm}|}
\hline 
Techniques & Cost Function Curve ($\rho(x)$) & Differentiation of the Cost Function: \\ Influence Curve ($\psi(x)$)\tabularnewline
\hline
\hline 
L1 Norm &  \raisebox{-1.8cm}{\includegraphics[width=2.7cm,height=1.9cm]{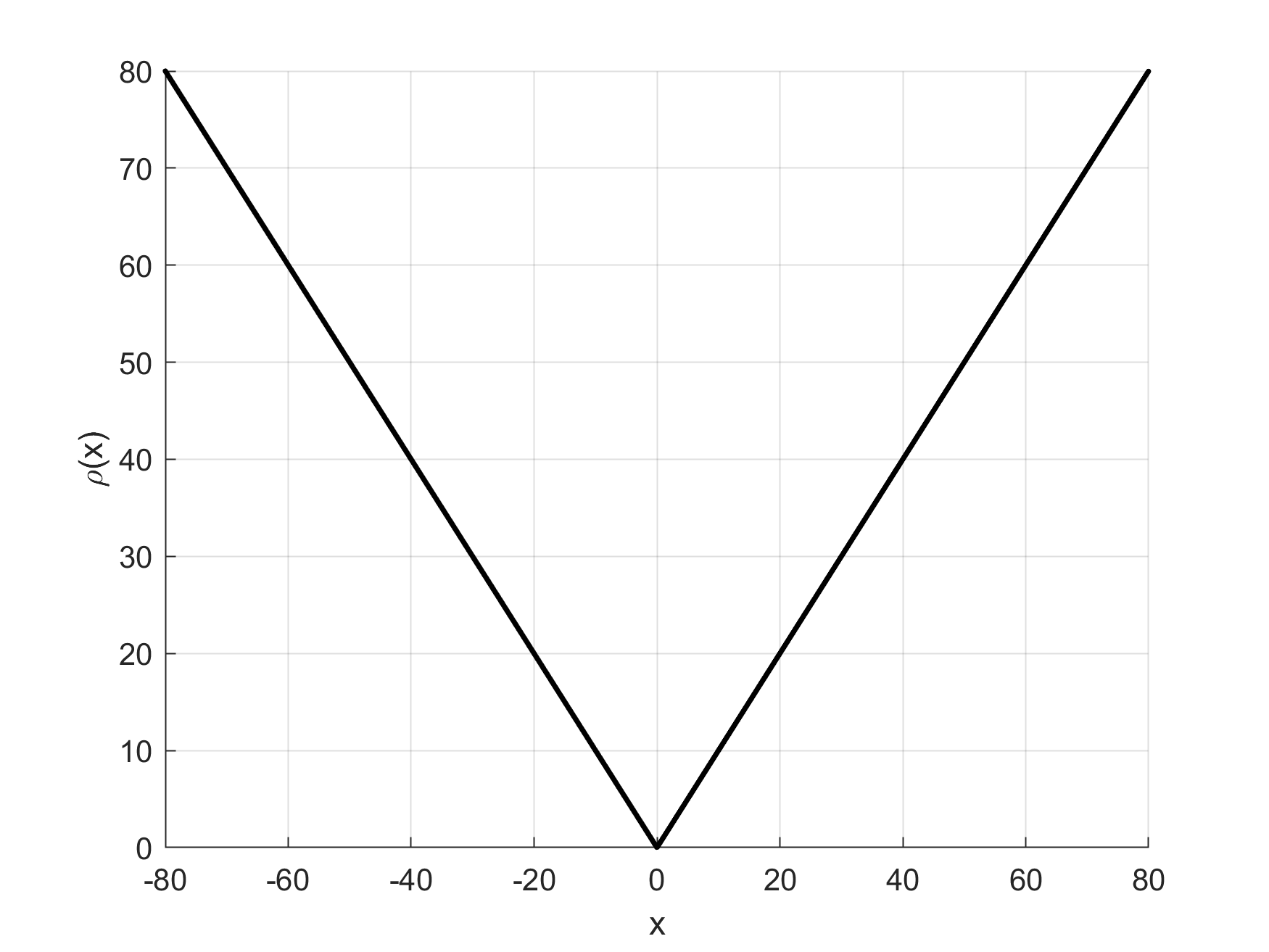}} & \raisebox{-1.8cm}{\includegraphics[width=2.7cm,height=1.9cm]{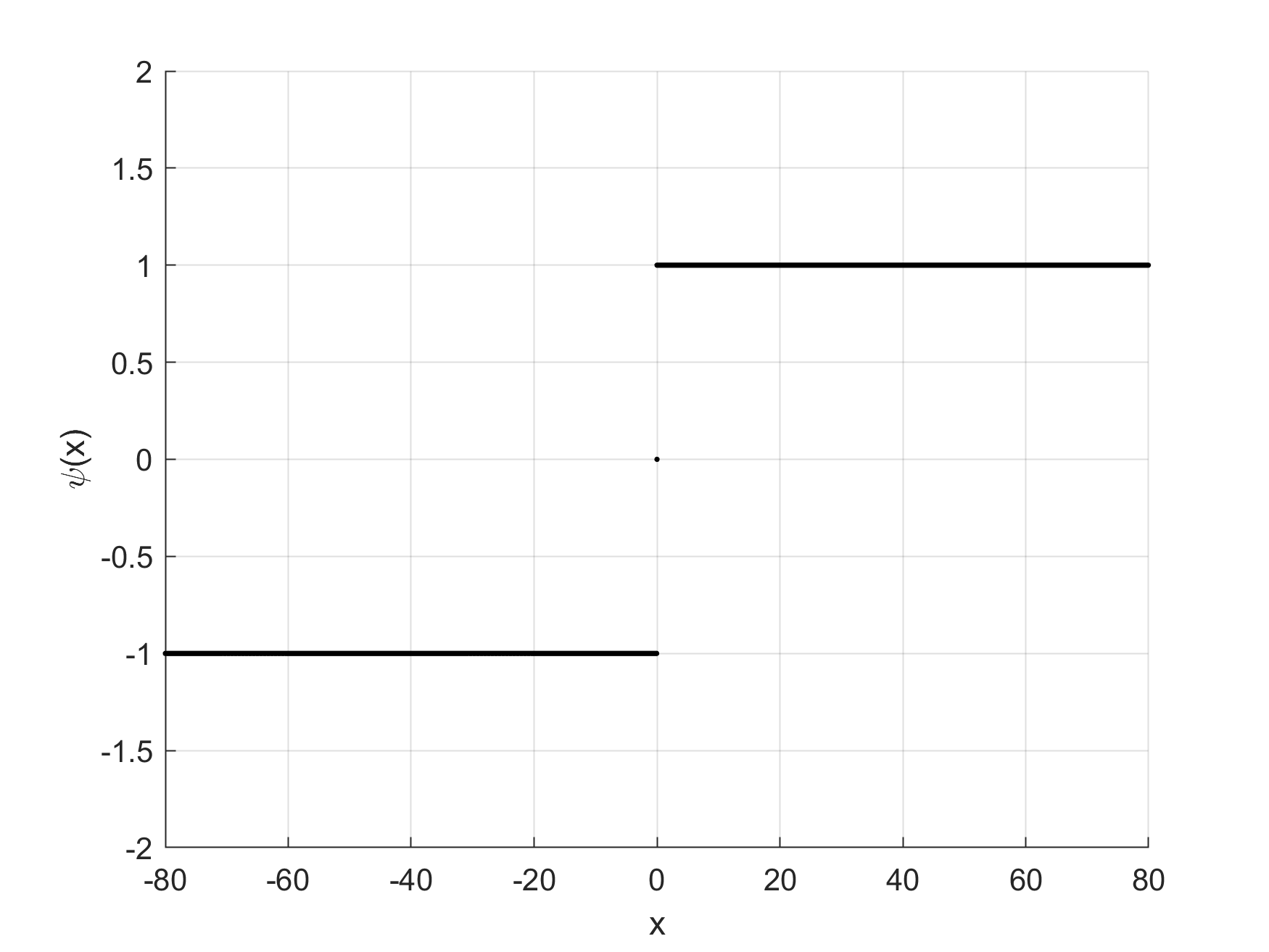}} \tabularnewline
\hline 
L2 Norm & \raisebox{-1.8cm}{\includegraphics[width=2.7cm,height=1.9cm]{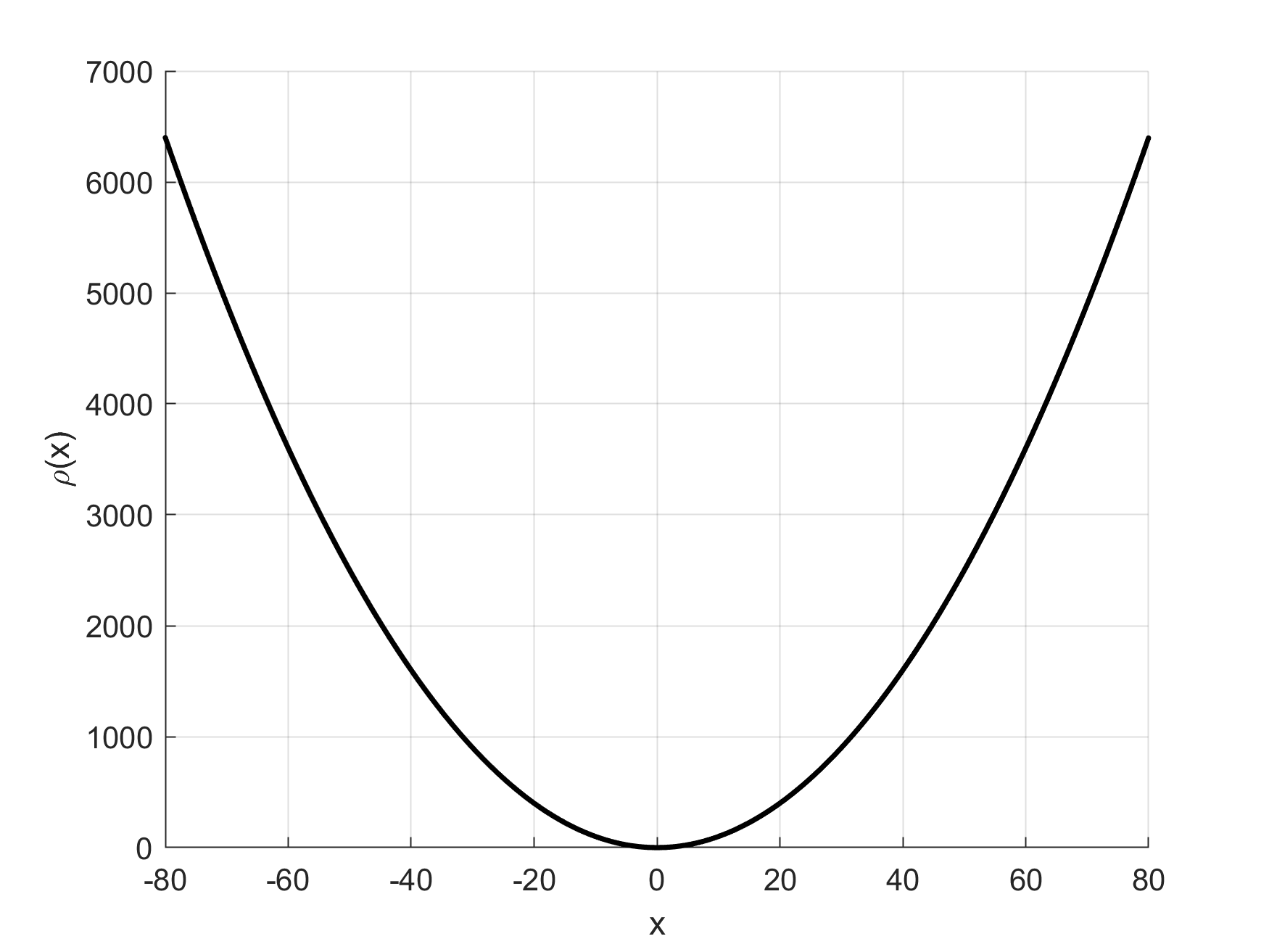}} & \raisebox{-1.8cm}{\includegraphics[width=2.7cm,height=1.9cm]{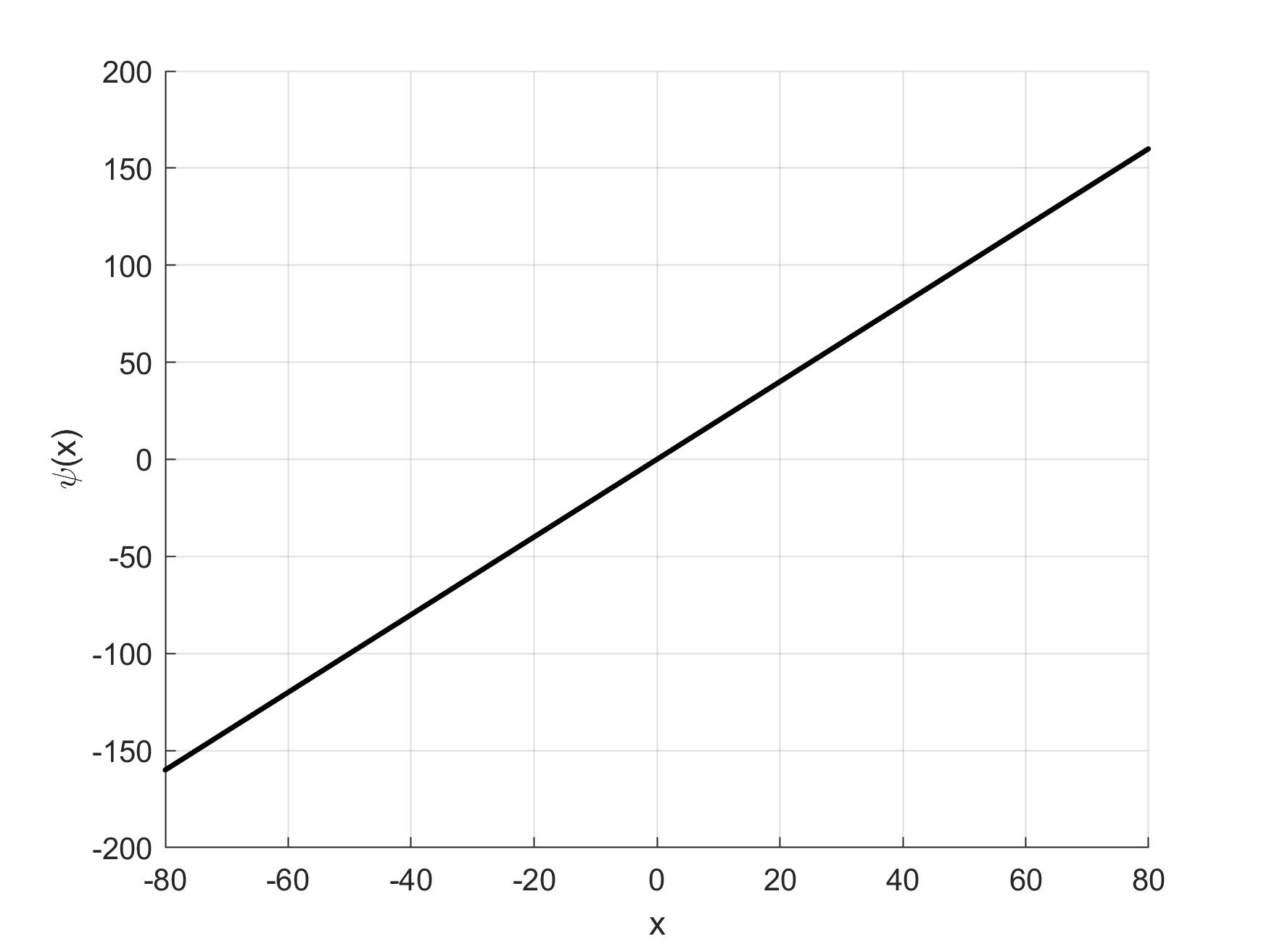}}\tabularnewline
\hline 
Huber's M-estimator ($k=20$) & \raisebox{-1.8cm}{\includegraphics[width=2.7cm,height=1.9cm]{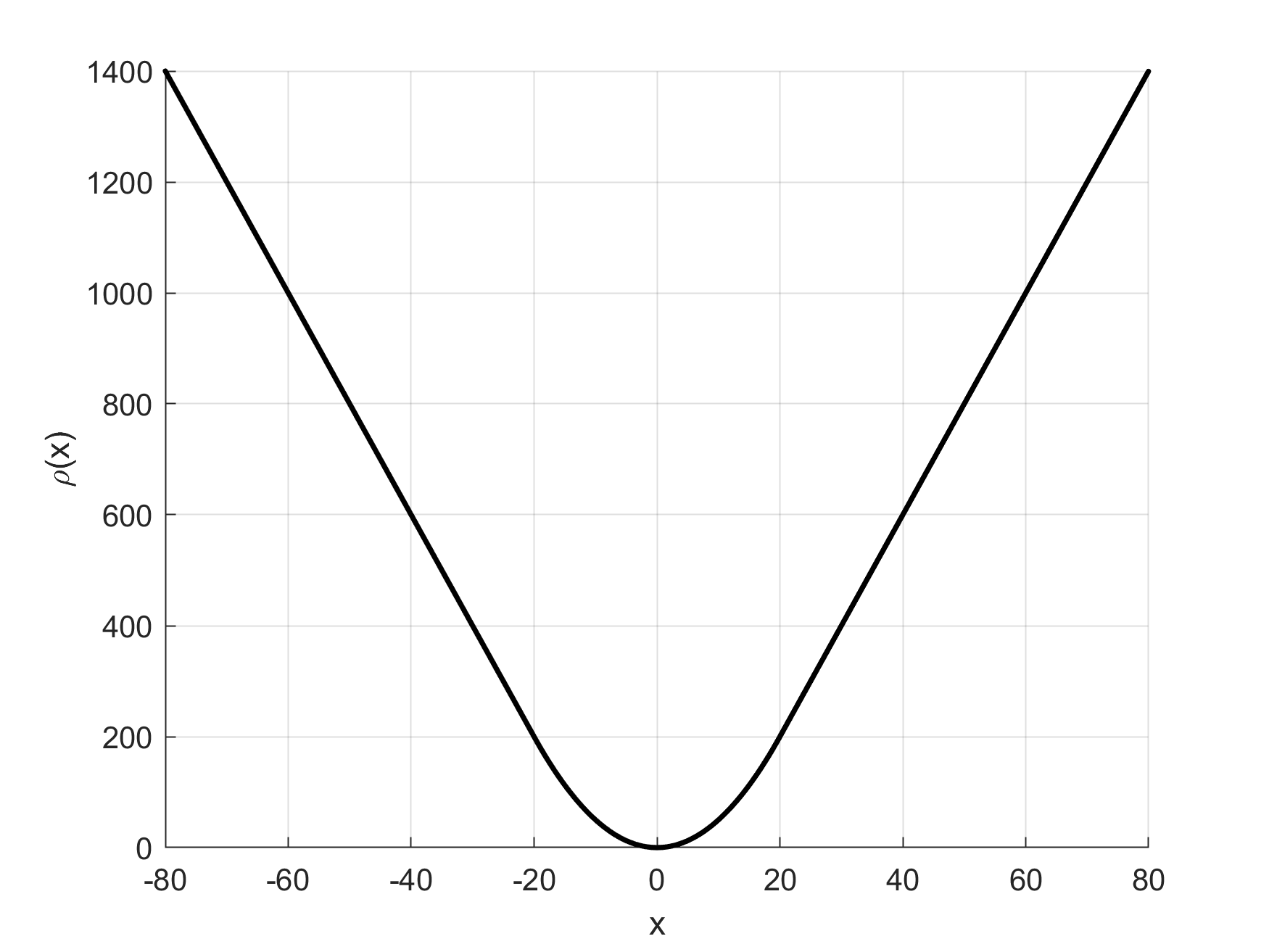}} & \raisebox{-1.8cm}{\includegraphics[width=2.7cm,height=1.9cm]{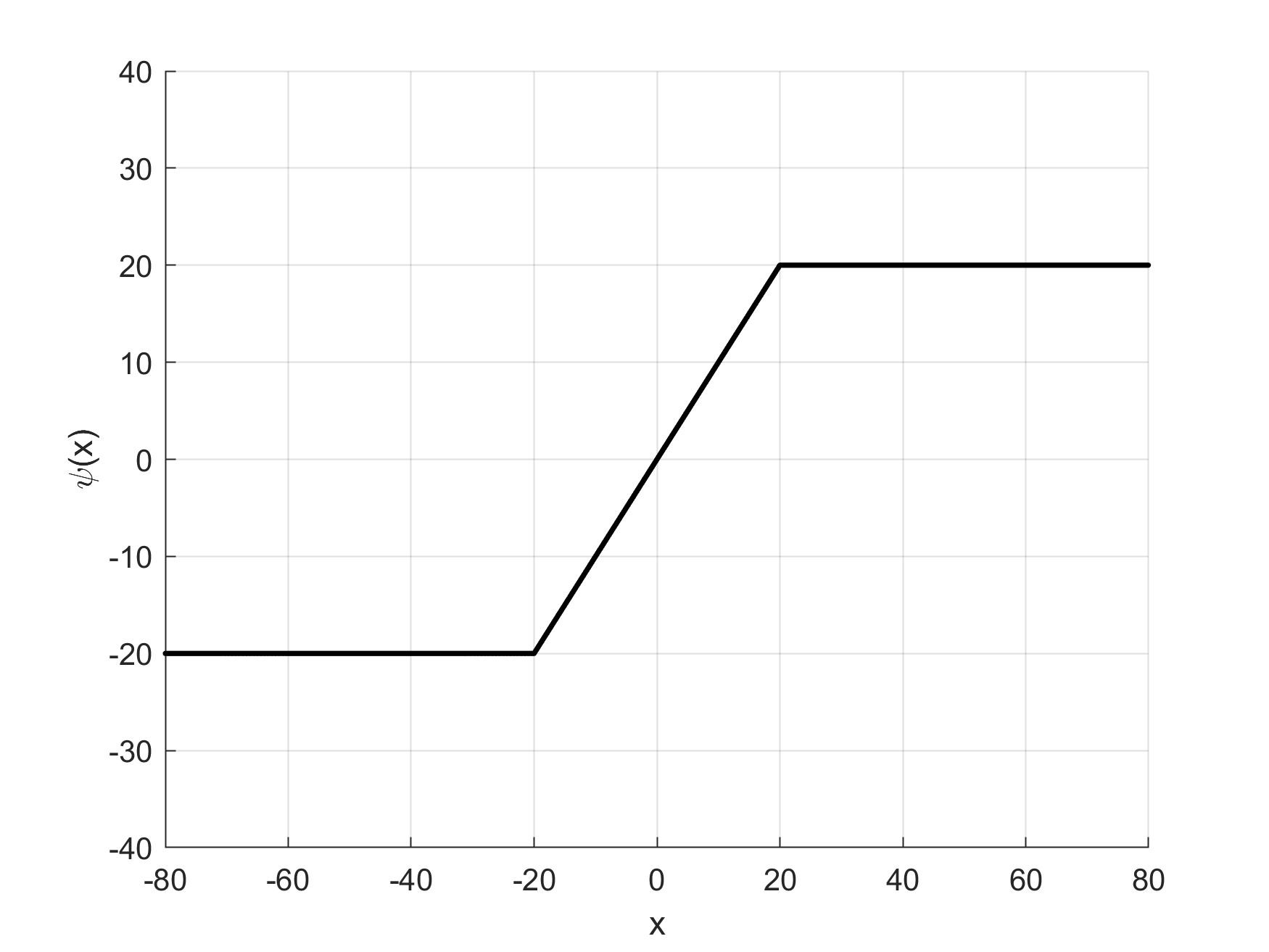}}\tabularnewline
\hline 
Tukey's M-estimator ($k=20$) & \raisebox{-1.8cm}{\includegraphics[width=2.7cm,height=1.9cm]{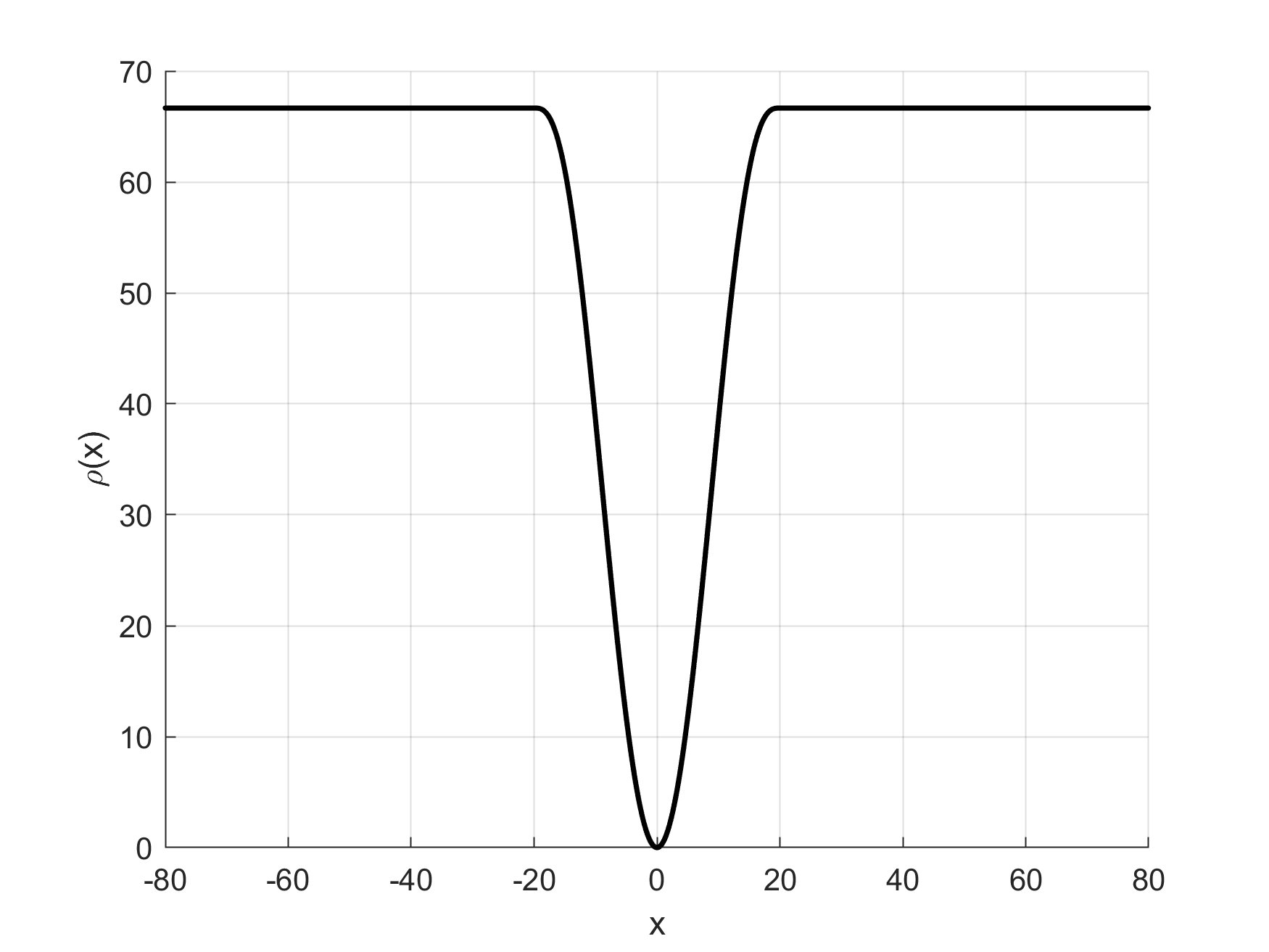}} & \raisebox{-1.8cm}{\includegraphics[width=2.7cm,height=1.9cm]{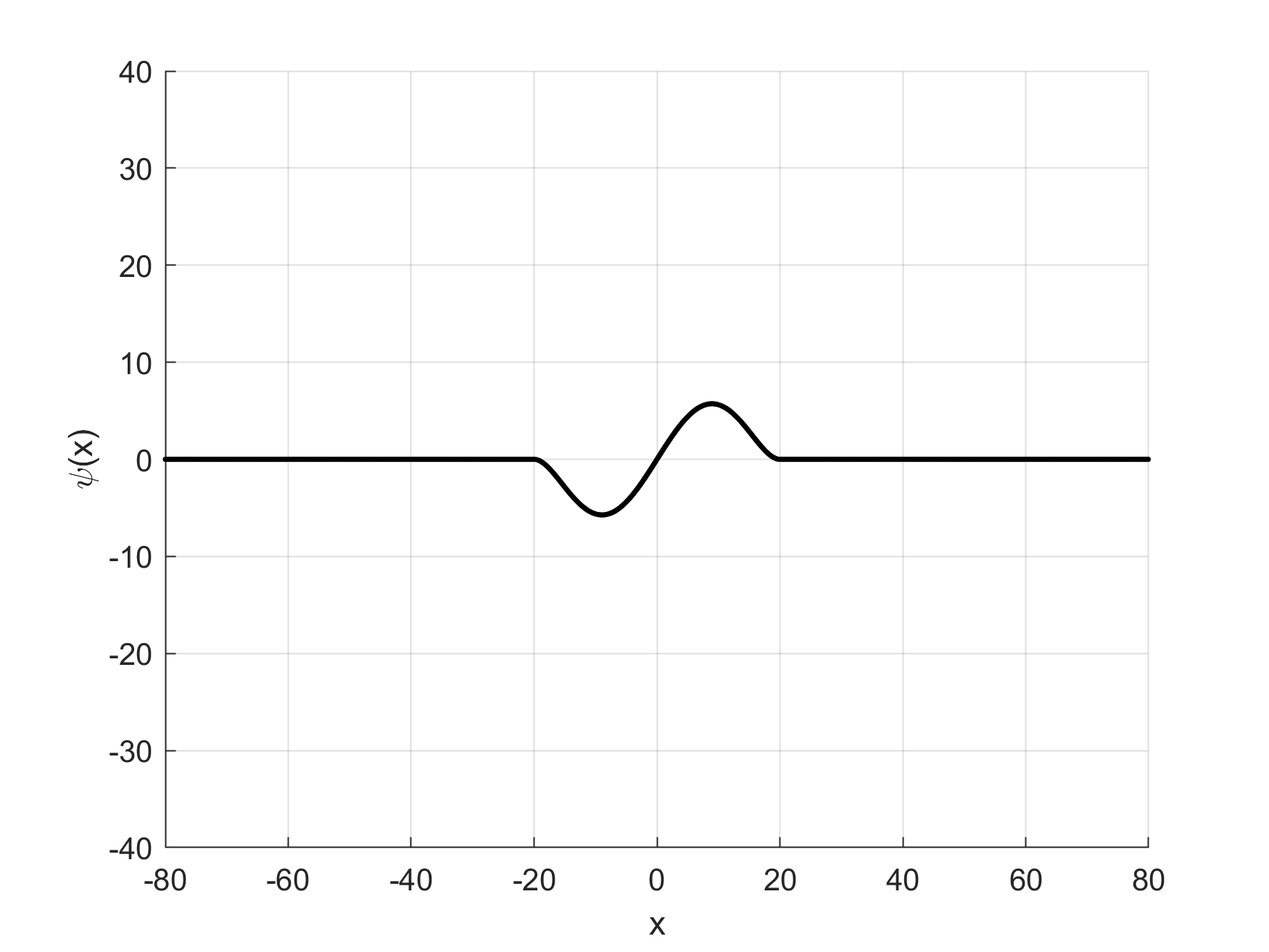}}\tabularnewline
\hline 
Cauchy-Lorentzian & \raisebox{-1.8cm}{\includegraphics[width=2.7cm,height=1.9cm]{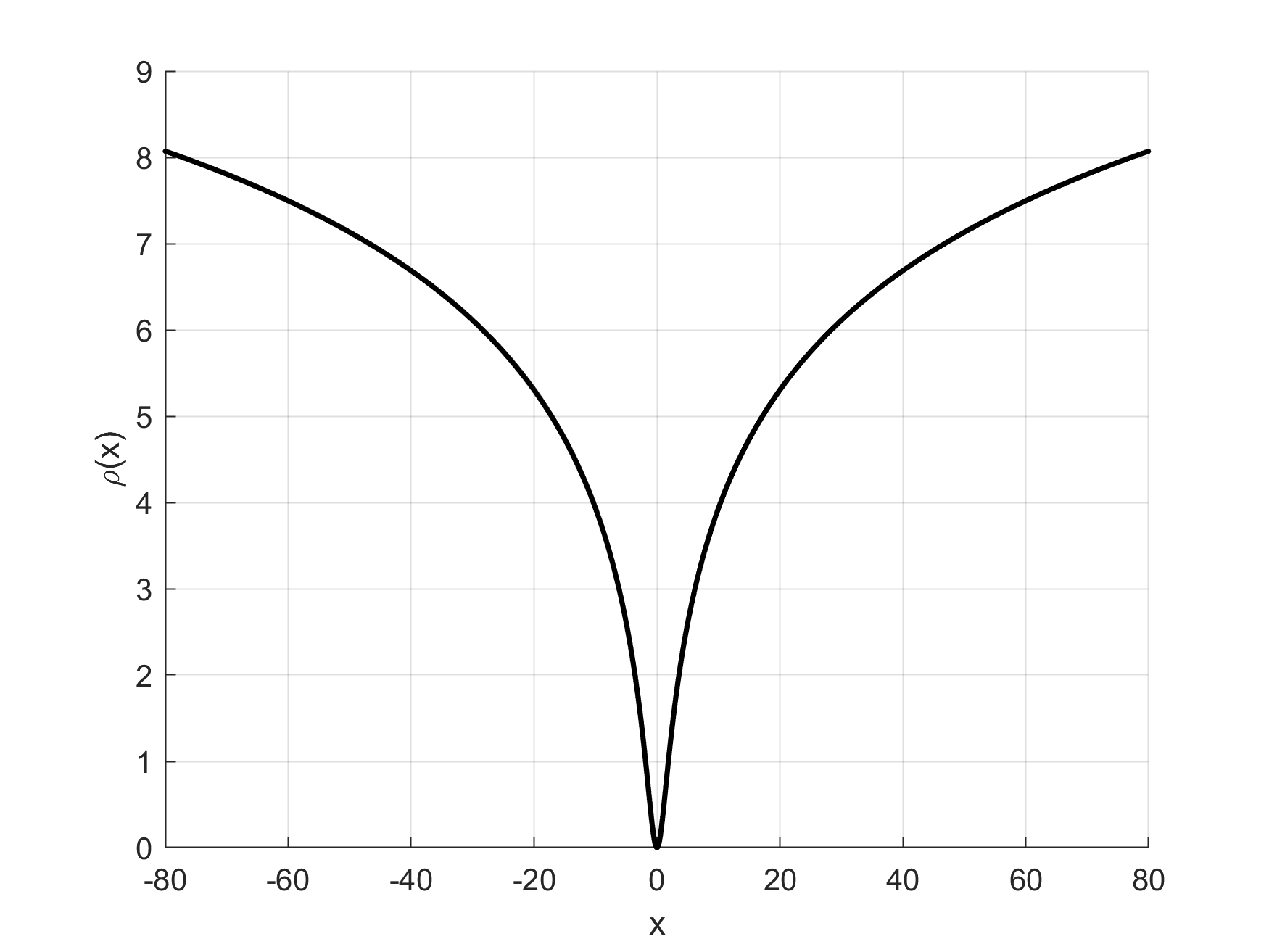}} & \raisebox{-1.8cm}{\includegraphics[width=2.7cm,height=1.9cm]{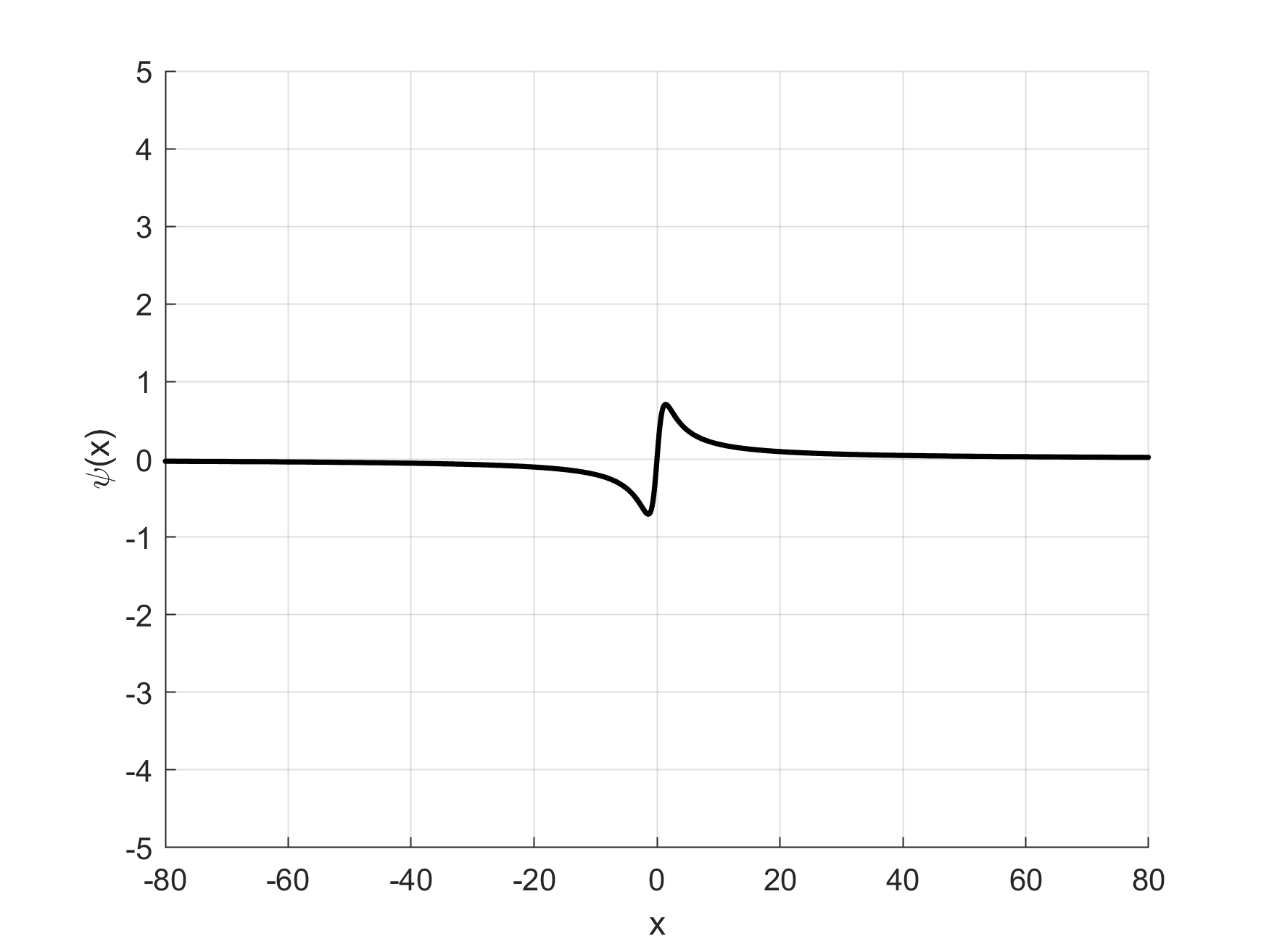}}\tabularnewline
\hline
Parameterised Student-t ($\tau$,$\nu$=20) & \raisebox{-1.8cm}{\includegraphics[width=2.7cm,height=1.9cm]{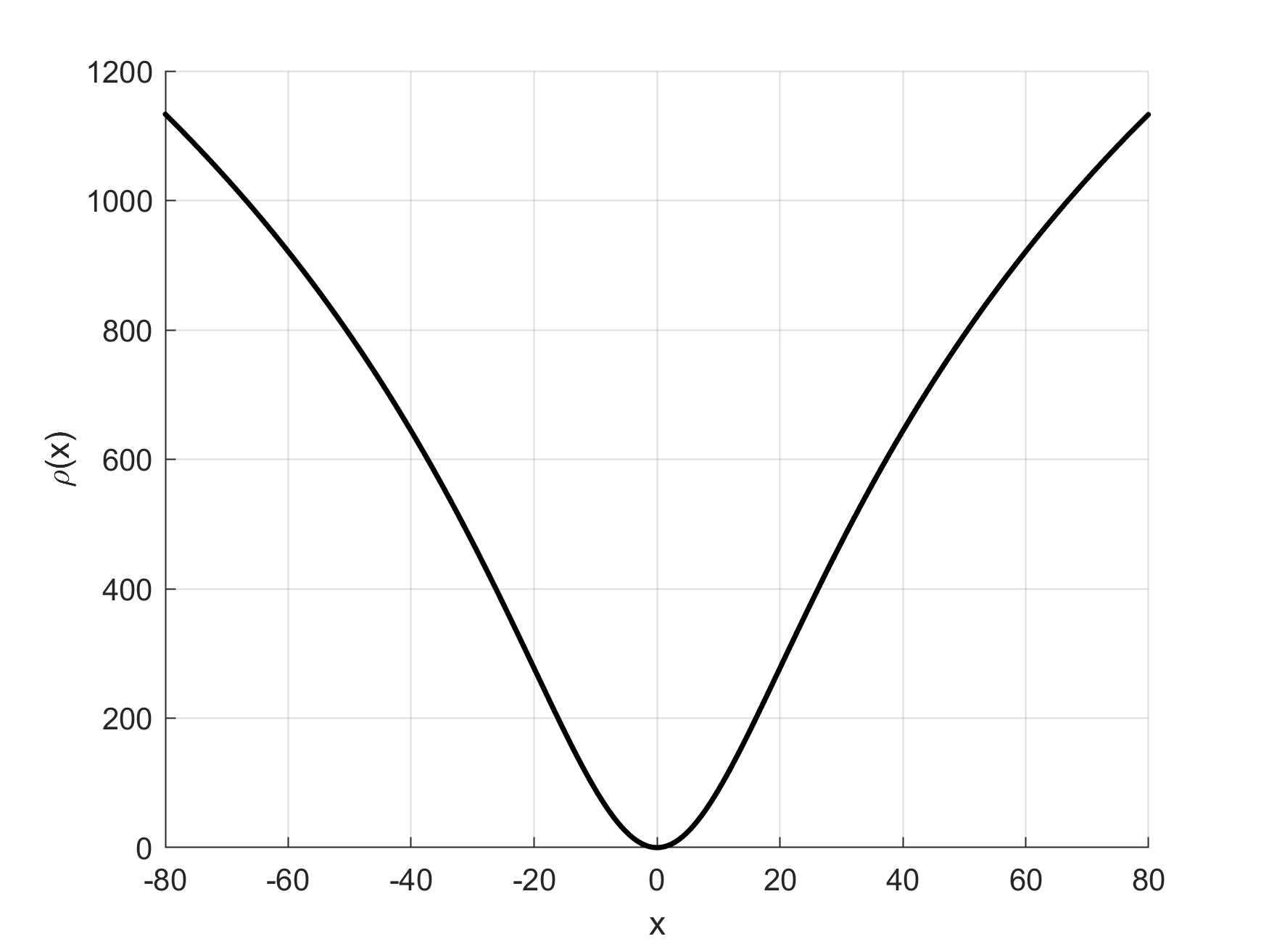}} & \raisebox{-1.8cm}{\includegraphics[width=2.7cm,height=1.9cm]{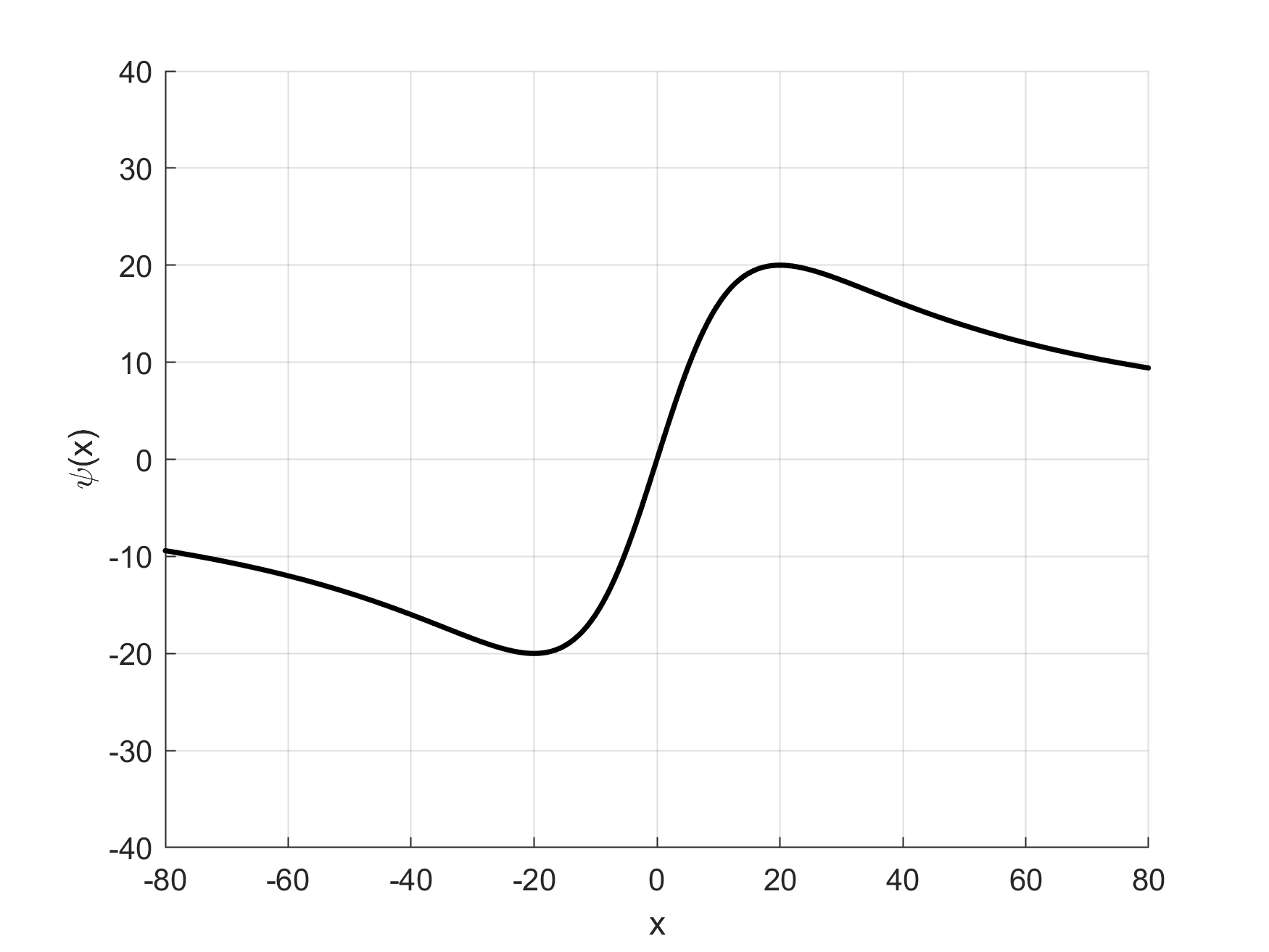}}\tabularnewline
\hline 
\end{tabular}
\par\end{centering}
\end{table}

\section{Parameterised Student-t Distribution}

\begin{figure}[htp]
\centering
\includegraphics[width=4.0cm,height=2.2cm]{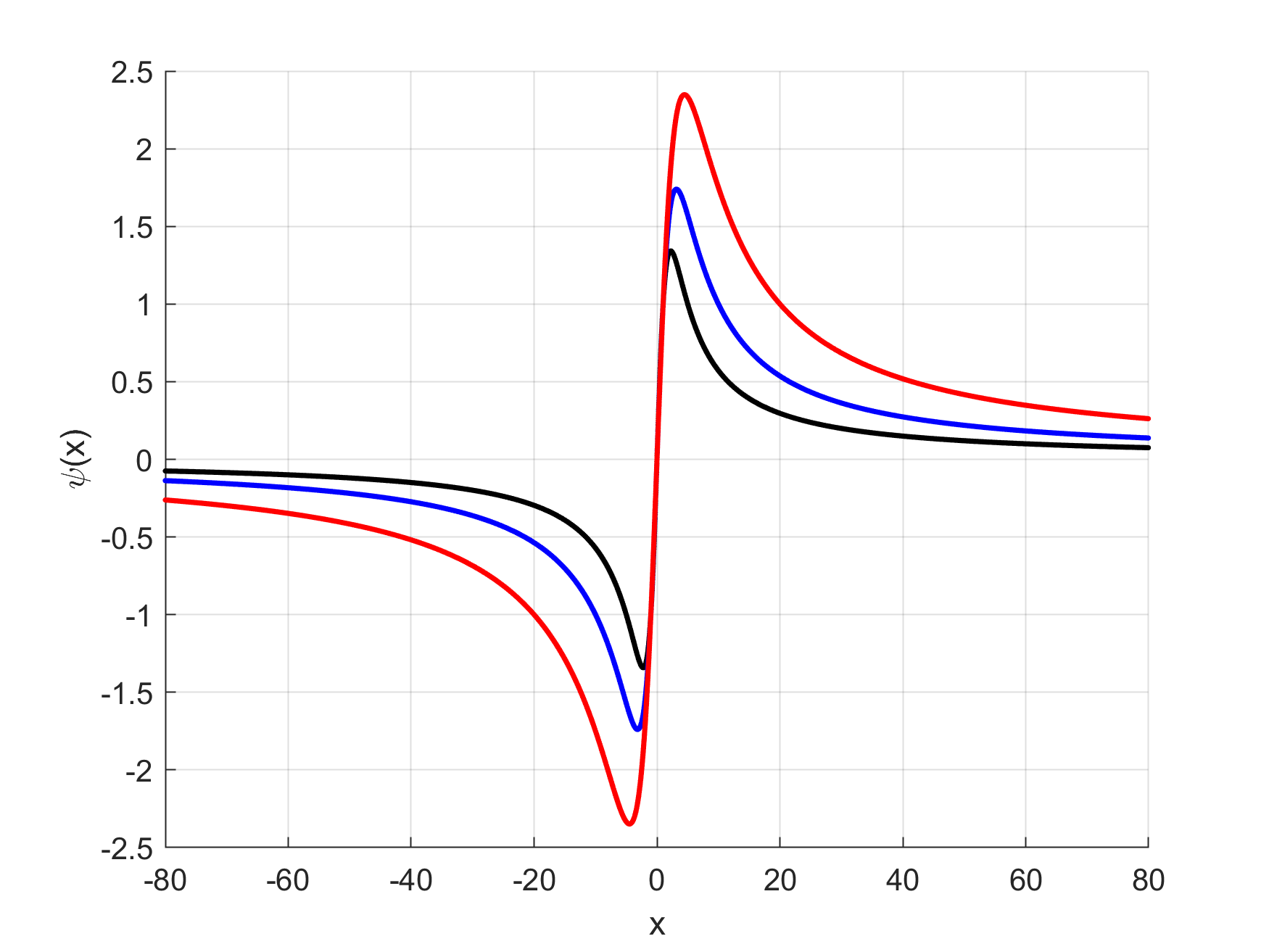}
\includegraphics[width=4.0cm,height=2.2cm]{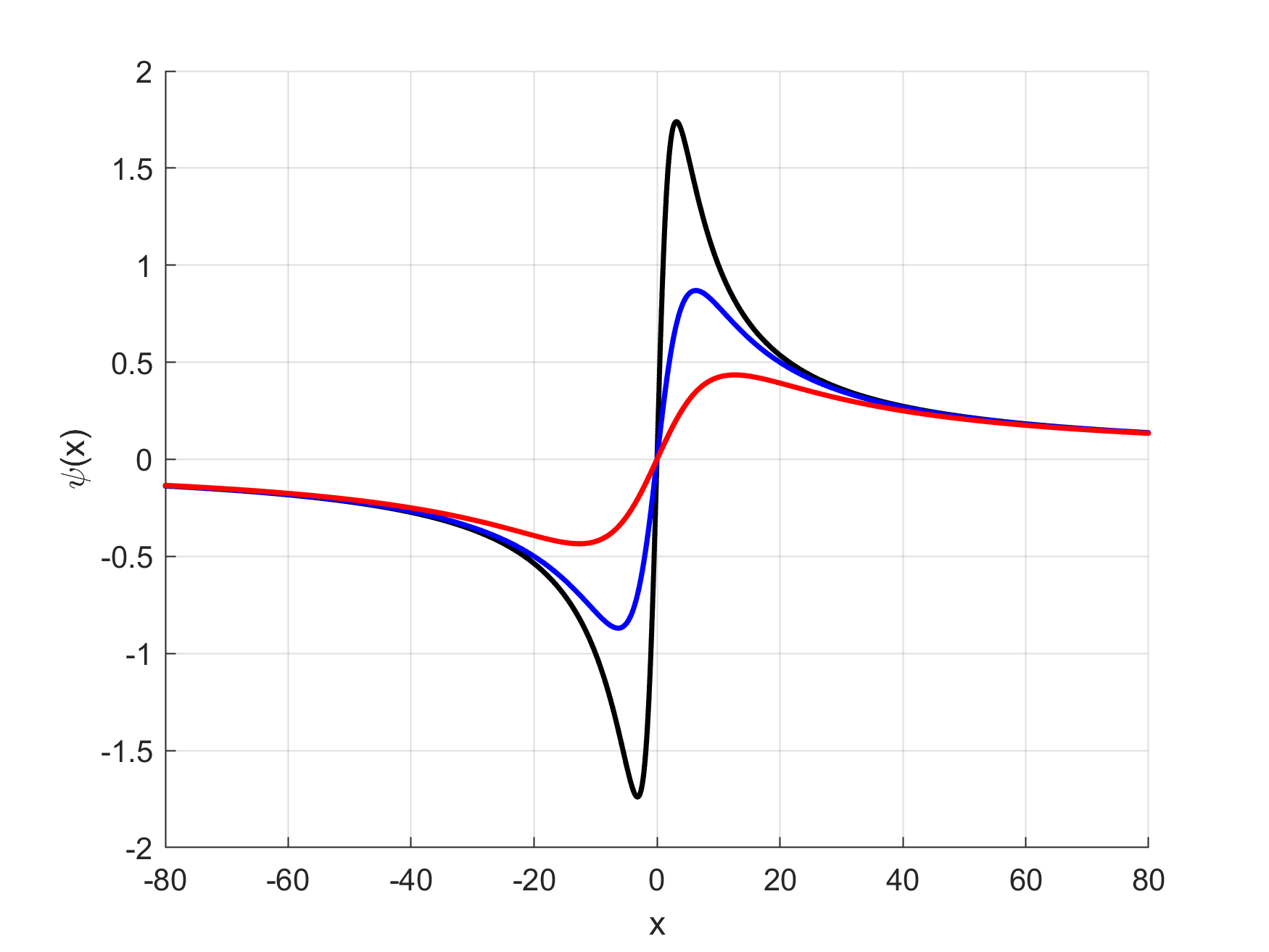}
\caption{The influence function ($\psi(x)$) of traditional student-t function as (2). (left) $\sigma = 1$, $v = 5, 10, 20$; (right) $\sigma = 1, 2, 4$, $v = 10$.\label{fig:old-st-curve}}
\end{figure}

\begin{figure}[htp]
\centering
\includegraphics[width=4.0cm,height=2.2cm]{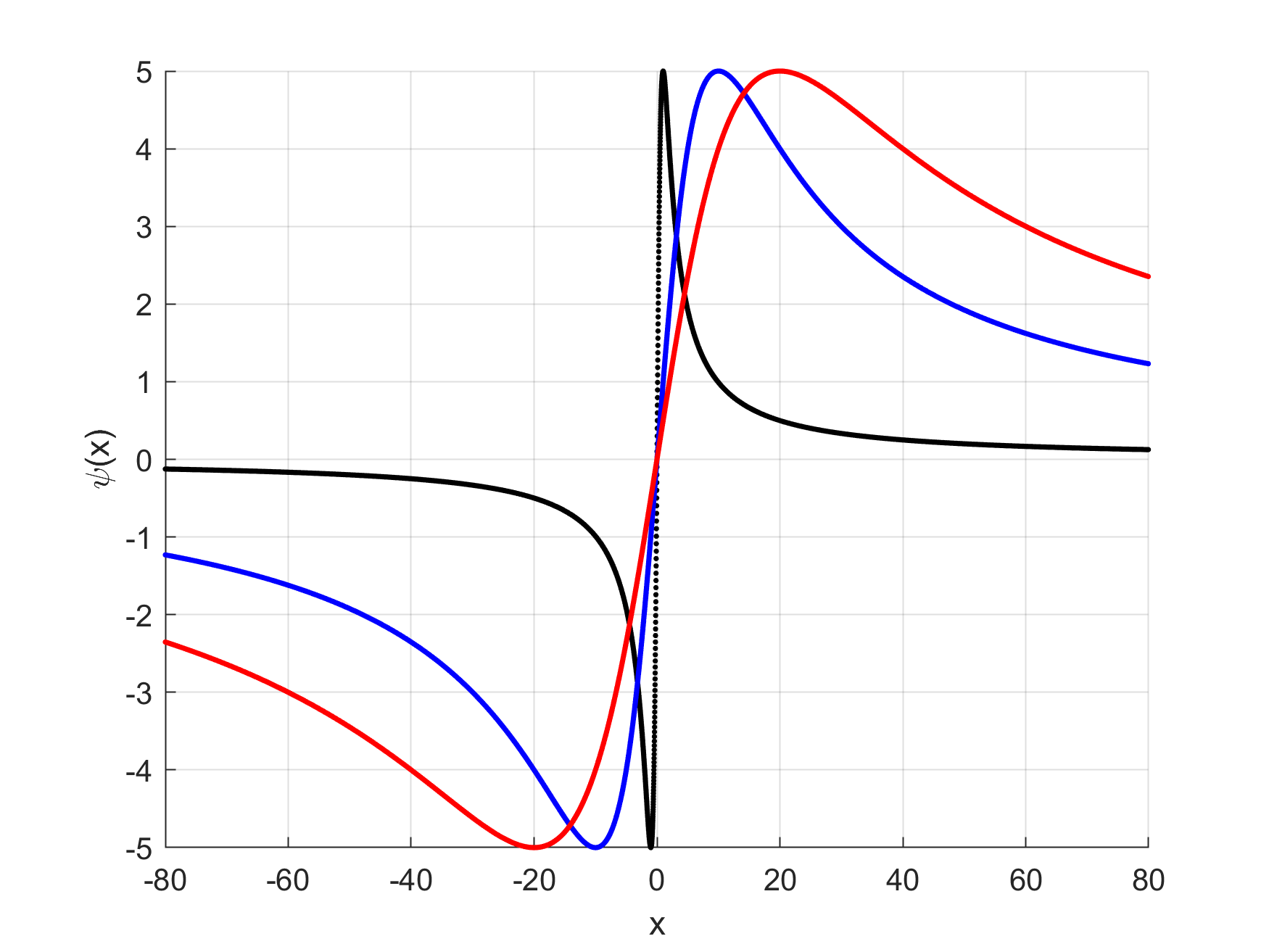}
\includegraphics[width=4.0cm,height=2.2cm]{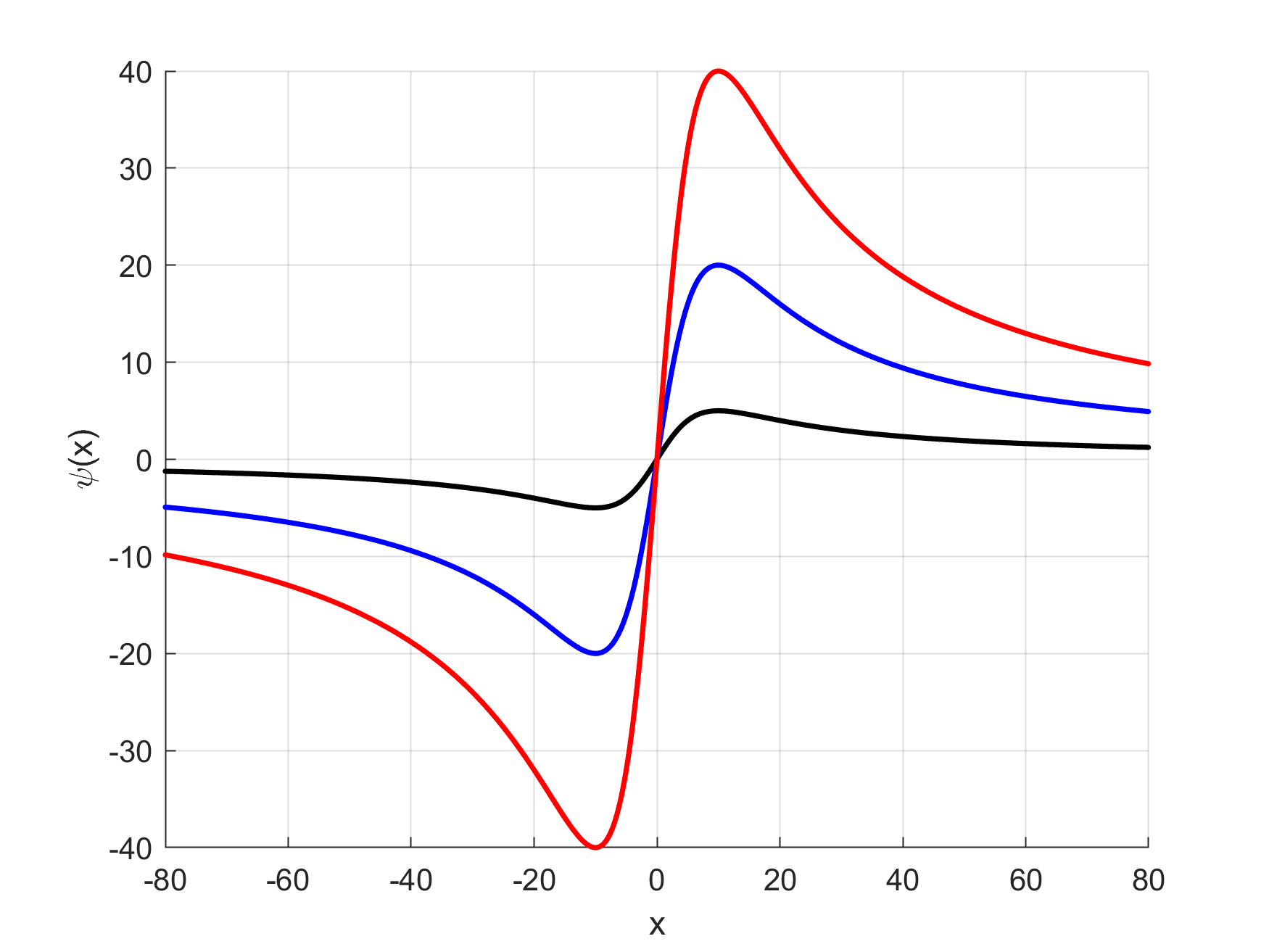}
\caption{The influence function ($\psi(x)$) of the parameterised student-t function as (4). (left) $\tau = 5$, $\nu = 1, 10, 20$; (right) $\tau = 5, 20, 40$, $\nu = 10$.\label{fig:new-st-curve}}
\end{figure}

It is not obvious how to choose a robust cost function that will facilitate efficient optimisation. Due to its ability to interpolate between the Gaussian distribution (when $v=\infty$) and the Cauchy-Lorentzian distribution (when $v=1$), the student t-distribution can be used in a way that adapts in response to each of a series of optimisation problems. In order to use it as a cost function, the student t-distribution is expressed in (\ref{eq:old-st-cost}) on a logarithmic scale. Note that such distribution is re-parameterised in a way that makes the interpolation explicit between the L2 norm and the Cauchy-Lorentzian function.

\begin{equation}
\label{eq:old-st-cost}
\rho(x)=\frac{v+1}{2}\ln(1+\frac{x^{2}}{v\sigma^{2}}) - \ln\left[\frac{\Gamma(\frac{v+1}{2})}{\sqrt{v\pi}\sigma\Gamma(\frac{v}{2})}\right]
\end{equation}

Figure~\ref{fig:old-st-curve} shows how the influence functions (the differentiation of (\ref{eq:old-st-cost})) change when only $v$ or only $\sigma$ changes. By considering the peak on the curve, it is obvious that neither $v$ nor $\sigma$ alone can be used to determine the x-coordinate of the peak's position or the peak's magnitude. When either $v$ or $\sigma$ increases, the peak becomes larger (vertically) and moves away from the origin. This makes it difficult to interpret the effect on the influence curve of changing each parameter: the x-position of the peak and the magnitude of the peak are not intuitively related to the parameters. Therefore, we propose the parameterisation of the student-t cost function and its influence function as follows.

\begin{equation}
\label{eq:new-st-cost}
\rho(x)=\tau\nu\ln(1+\frac{x^{2}}{\nu^{2}})-\ln\left[\frac{\Gamma(\tau\nu)}{\sqrt{\pi}\nu\Gamma(\tau\nu-\frac{1}{2})}\right]
\end{equation}

\begin{equation}
\label{eq:new-st-influ}
\psi(x)=\frac{2\tau\nu x}{\nu^{2}+x^{2}}
\end{equation}

The re-parameterisation replaces $v$ and $\sigma$ by $\nu$ and $\tau$. Note that this incurs no additional computational cost. Figure~\ref{fig:new-st-curve} shows how the cost function changes when modifying $\nu$ and $\tau$. It should be clear to the reader that the position (on the $x$-axis) of the peak equals $\nu$ and the magnitude of the peak (on the $y$-axis) equals $\tau$. The re-parameterisation provides a clear interpretation in terms of the parameters' effect on the influence curve. In the context of optimisation, a larger $\nu$ means that high-error samples have a greater influence. Such a larger $\nu$ results in an estimator that is less robust. However, since more samples are involved in the calculation, especially since the larger errors are included, the optimisation should be quicker and more robust to poor initialisation. Thus, the optimal current compromise between efficiency, ability to respond to poor initialisation and robustness becomes difficult to specify a priori and is therefore well suited to online adaption. In contrast, the other parameter $\tau$ decides the maximum influence of a datum. This parameter behaves like a learning rate in an iterative algorithm and will not be our focus. In this letter, it is recommended that $\tau=\nu$, which makes the curve similar to the L2 norm for errors with a magnitude less than $\nu$.

\section{Robust Global Motion Estimation}

\subsection{Optimisation}

Global motion estimation aims to find a set of motion parameters that warp the input image. We consider an illustrative application where the altitude of the airborne camera is sufficient that an affine transformation can be considered. Recall the affine transformation model is as described in (\ref{eq:motion-model}):

\begin{equation}
\label{eq:motion-model}
\left[\begin{array}{cc}
x & y
\end{array}\right]^{T}=f(x',y';A)=A\cdotp\left[\begin{array}{ccc}
x' & y' & 1\end{array}\right]^{T}
\end{equation} 

\noindent where

$A=\left[\begin{array}{ccc}
a_{xx} & a_{xy} & a_{x}\\
a_{yx} & a_{yy} & a_{y}
\end{array}\right]$ is the transformation matrix containing motion parameters to be calculated. $\begin{bmatrix}x &  y\end{bmatrix}$ are the warped coordinates, and $\begin{bmatrix}x' &  y'\end{bmatrix}$ are the original coordinates.

The GME problem can be formulated as (\ref{eq:error}) with the student-t cost function. The total error $\varepsilon$ should be minimised by estimating $A$:

\begin{eqnarray}
\label{eq:error}\label{eq:img-cost-func}
\varepsilon & = & \sum_{x=1}^{\omega_{x}}\sum_{y=1}^{\omega_{y}}\tau\nu\ln(1+\frac{\left (I_{ref}^{(x,y)}-I_{reg}^{(x,y)}\right)^{2}}{\nu^{2}})\nonumber \\
 &  & -\ln\left[\frac{\Gamma(\tau\nu)}{\sqrt{\pi}\nu\Gamma(\tau\nu-\frac{1}{2})}\right]
\end{eqnarray}

\begin{equation}
I_{reg} = F(I_{ref}, A);
\end{equation}

\noindent where $F(.)$ is the warping process and image interpolation may be applied during the warping if necessary. $I_{reg}$ is the warped/registered image. $I_{ref}$ is the reference/original image.

Newton's method is used to solve the optimisation problem, and the first-order derivative of~(\ref{eq:img-cost-func}) is (\ref{eq:first-order-err}):

\begin{equation}
\label{eq:first-order-err}
\frac{\partial\varepsilon}{\partial A} = \sum_{x=1}^{\omega_{x}}\sum_{y=1}^{\omega_{y}}\frac{-2\tau\nu\left (I_{ref}^{(x,y)}-I_{reg}^{(x,y)}\right)}{\nu^{2}+\left(I_{ref}^{(x,y)}-I_{reg}^{(x,y)}\right)^{2}}\cdotp\vartheta(x,y)
\end{equation}

\begin{equation}
\vartheta(x,y) = \frac{\partial I_{reg}^{(x,y)}}{\partial A}
\end{equation}

\noindent where $\omega_{x} \times \omega_{y}$ is the resolution of the reference image.

The second-order derivative of~(\ref{eq:img-cost-func}) is:

\begin{eqnarray}
\label{eq:second-order-err}
\frac{\partial^{2}\varepsilon}{\partial A^{2}} & = & \sum_{x=1}^{\omega_{x}}\sum_{y=1}^{\omega_{y}}\left(\frac{2\tau\nu(\nu^{2}-I_{diff}^{2})}{\nu^{2}+I_{diff}^{2}}\cdotp\vartheta(x,y)^{T}\cdotp\vartheta(x,y)\right.\nonumber \\
 &  & +\left. \underbrace{ \frac{2\tau\nu I_{diff}}{\nu^{2}+I_{diff}^{2}}\cdotp\frac{\partial}{\partial A^{'}}\vartheta(x,y)}_{\approx 0}\right)
\end{eqnarray}

\begin{equation}
I_{diff}=I_{ref}^{(x,y)}-I_{reg}^{(x,y)}
\end{equation}

We approximate the second derivative term to zero to reduce computational cost in (\ref{eq:second-order-err}). To calculate $\vartheta$, we consider:

\begin{equation}
\label{eq:theta1}
\vartheta(x,y)=\frac{\partial I_{reg}^{(x,y)}}{\partial\left[
x,y\right]^{T}}\cdotp\frac{\partial f(x',y';A)}{\partial A}=\nabla I_{reg}^{(x,y)}\cdotp\frac{\partial f(x,y;A)}{\partial A}
\end{equation}

\noindent where $\nabla I_{reg}^{(x,y)}$ is the image gradient $[\frac{\partial I_{reg}^{(x,y)}}{\partial x}, \frac{\partial I_{reg}^{(x,y)}}{\partial y}]^{T}$ relative to x-axis and y-axis and:

\begin{equation}
\frac{\partial f(x',y';A)}{\partial A}=\left[\begin{array}{cccccc}
x & y & 1 & 0 & 0 & 0\\
0 & 0 & 0 & x & y & 1
\end{array}\right] 
\end{equation}

\noindent which comes from (\ref{eq:motion-model}).

Therefore, (\ref{eq:theta2}) is used in the optimisation.

\begin{equation}
\label{eq:theta2}
\vartheta=[\begin{array}{cccccc}
x\nabla I_{est}^{x} & y\nabla I_{est}^{x} & \nabla I_{est}^{x} & x\nabla I_{est}^{y} & y\nabla I_{est}^{y} & \nabla I_{est}^{y}\end{array}]
\end{equation}

According to Newton's method, the transformation matrix $A$ is updated iteratively via (\ref{eq:motion-model-update}) and initialised with (\ref{eq:motion-model-init}). Moreover, a learning rate $\lambda$ is involved, which is usually 1. In practice, we may need to use a smaller learning rate when the reference image and input image are highly inconsistent to avoid large estimation errors. In~\cite{bouguet2001pyramidal}, to reduce the computational cost, $\nabla I_{ref}$ is used to alter $\nabla I_{reg}$, which is only calculated once before the iterative estimation. Note that the coarse-to-fine technique, which is usually used in image registration, is also implemented.

\begin{equation}
\label{eq:motion-model-update}
A_{est}=A_{pre}-\lambda\cdotp(\frac{\partial^{2}\varepsilon}{\partial A^{2}})^{-1}\cdotp\frac{\partial\varepsilon}{\partial A}
\end{equation}

\begin{equation}
\label{eq:motion-model-init}
A_{init} = \left[\begin{array}{ccc}
1 & 0 & 0\\
0 & 1 & 0
\end{array}\right]
\end{equation}

\subsection{Semi-offline Parameter Estimation}\label{sec:ParamGen}

The involvement of $\tau$ and $\nu$ is the biggest advantage of the proposed parameterised student-t function, and its value is significant. Note that, in Newton's method, $\tau$ is eliminated (see (\ref{eq:first-order-err}), (\ref{eq:second-order-err}) and (\ref{eq:motion-model-update})). In order to find the best $\nu$ that trades off the accuracy and time cost, a binary search-like scenario is adopted (see Algorithm~\ref{alg:chap2-st-gme}). Since a sequence of images tends to have similar content in a small period, $\nu$ does not need to be estimated often. In our experiment, Algorithm~\ref{alg:chap2-st-gme} is only activated at the beginning of a sequence or when the global error, $\sum_{x=1}^{\omega_{x}}\sum_{y=1}^{\omega_{y}}(I_{ref}^{(x,y)}-I_{reg}^{(x,y)})$, changes considerably compared to that achieved when processing the previous image.

\begin{algorithm}
\caption{Parameter estimation scenario.}\label{alg:chap2-st-gme}
\begin{algorithmic}[1]
\State Perform global motion estimation with two potential parameters $\nu_{max}$ and $\nu_{min}$. The errors are $\varepsilon_{max}$ and $\varepsilon_{min}$. $cnt = 0$. $T_{err}$ and $T_{cnt}$ are two pre-defined variables.

\While {$|\varepsilon_{max}-\varepsilon_{min}| > T_{err}$ or $cnt < T_{cnt}$}

\State Keep $\nu_{max}$ or $\nu_{min}$ whichever has the smaller error. The other one will be replaced by $\frac{\nu_{max}+\nu_{min}}{2}$.
\State Perform global motion estimation again with the parameters $\nu_{max}$ and $\nu_{min}$ and calculate the errors, $\varepsilon_{max}$ and $\varepsilon_{min}$.
\State $cnt=cnt+1$

\EndWhile

\State Choose from $\nu_{max}$ and $\nu_{min}$ which produces smaller GME error for the subsequent optimisations.

\end{algorithmic}
\end{algorithm}

\section{Experiment Results}

The experiments are conducted on four datasets in the Vivid benchmark~\cite{collins2005open}. The resolution of the videos is $640 \times 480$. The images contain noise and sometimes are blurred. The experiments compare the L1 norm, L2 norm, Huber's and Tukey's M-estimators, and the proposed student-t cost function. For Huber's and Tukey's M-estimator, the only parameter is pre-defined: $k=20$ (for the detailed definitions, we refer to~\cite{fox2002robust}). For the proposed student-t cost function, we use two variants. One (`Stu-t1') involves pre-defining the parameters as: $\tau=\nu=20$. Such that the three cost functions are similar: $\psi(x)$ is limited when $|x|\geq20$ (the curves are shown in Table~\ref{tab:curves}). The other variant (`Stu-t2') involves using the proposed parameter adaptation method (in Section~\ref{sec:ParamGen}) with $\tau_{max=40}$ and $\tau_{min}=10$. To create comprehensive results, we reduce the frame rate to approximately 8~Hz and 4~Hz. There will be a large displacement between two images, such that the optimisation processes can easily become stuck in local minima. Moreover, a contribution mask that considers the middle $610 \times 450$ area in each image, is used.

In terms of evaluation, because the mean squared error is easily dominated by outliers, we use an extended L0 norm:

\begin{equation}
\label{eq:evaluation}
E=\sum_{x=1}^{\omega_{x}}\sum_{y=1}^{\omega_{y}}\varepsilon^{(x,y)}
\end{equation}

\[
\varepsilon^{(x,y)}=\begin{cases}
1 & \qquad if|I^{(x,y)}-J^{(x,y)}|>c\\
0 & \qquad if|I^{(x,y)}-J^{(x,y)}|\leq c
\end{cases}
\]

\noindent where $I^{(x,y)}$ and $J^{(x,y)}$ are pixels on two images. When $c=0$, this is the L0-norm. We use $c=2$ in the experiments.

Table~\ref{tab:exp-mean-err} shows the average extended L0 norm errors from using different cost functions over four datasets. The performances of the cost functions show significant differences when the datasets and frame rates are varied: except for the L2 norm, all the cost functions can yield satisfying estimations (`satisfying' means a small difference, e.g., 0.2 [$\times 10^4$], compared to the best) on some datasets. It is clear that the proposed student-t cost function (with fixed parameters, shown in the column `Stu-t1') can produce 6 satisfying results. Although it is never the best one, it does not fall behind. When applying the proposed method for adapting parameters, this cost function (column `Stu-t2') performs the best in five out of eight experiments. In the other three experiments, the difference between this approach and the best approach is smaller than 0.1~[$\times 10^4$].

Table~\ref{tab:exp-mean-iter-num} shows the average numbers of iterations to converge using different cost functions. This table describes the computational cost of using the cost functions. By inspecting the figures under `Stu-t2' and `Tukey' in the rows `Viv1(8Hz)', `Viv3(8Hz)' and `Viv3(4Hz)', we can see the trade-off between the computational cost and the accuracy.

\begin{table}
\caption{The mean errors [$\times10^4$] of the estimations over all pairs of images. Sampling Rates are shown in the brackets. The bold figures are the smallest errors. \label{tab:exp-mean-err}}
\begin{centering}
\begin{tabular}{|>{\centering}m{1.3cm}|>{\centering}m{0.7cm}|>{\centering}m{0.7cm}|>{\centering}m{0.7cm}|>{\centering}m{0.7cm}|>{\centering}m{0.7cm}|>{\centering}m{0.7cm}|}
\hline 
Dataset & L1 & L2 & Huber & Tukey & Stu-t1 & Stu-t2\tabularnewline
\hline
\hline 
Viv1(8Hz) & 6.75 & 8.97 & 6.94 & \textbf{5.20} & 5.56 & 5.27 \tabularnewline
\hline 
Viv1(4Hz) & 11.16 & 11.04 & 7.53 & 5.93 & 5.92 & \textbf{5.82} \tabularnewline
\hline 
Viv2(8Hz) & 4.87 & 5.84 & 5.59 & 5.34 & 5.07 & \textbf{4.84} \tabularnewline
\hline 
Viv2(4Hz) & 6.16 & 7.29 & 6.28 & 5.78 & 5.40 & \textbf{5.26} \tabularnewline
\hline 
Viv3(8Hz) &12.59 & 12.26 & 12.22 & \textbf{12.05} & 12.15 & 12.08 \tabularnewline
\hline
Viv3(4Hz) & 14.01 & 13.23 & 13.01 & \textbf{12.64} & 12.80 & 12.68 \tabularnewline
\hline 
Viv4(8Hz) & 11.80 & 9.60 & 9.05 & 9.44 & 8.83 & \textbf{8.77} \tabularnewline
\hline 
Viv4(4Hz) & 18.95 & 11.85 & 11.09 & 11.36 & 10.79 & \textbf{10.70} \tabularnewline
\hline 
\end{tabular}
\par\end{centering}
\end{table}

\begin{table}
\caption{The mean iteration numbers for the optimisation process. The bold figures correspond to the smallest errors in Table~\ref{tab:exp-mean-err}. \label{tab:exp-mean-iter-num}}
\begin{centering}
\begin{tabular}{|>{\centering}m{1.3cm}|>{\centering}m{0.7cm}|>{\centering}m{0.7cm}|>{\centering}m{0.7cm}|>{\centering}m{0.7cm}|>{\centering}m{0.7cm}|>{\centering}m{0.7cm}|}
\hline 
Dataset & L1 & L2 & Huber & Tukey & Stu-t1 & Stu-t2\tabularnewline
\hline
\hline 
Viv1(8Hz) & 26.5 & 10.7 & 8.8 & \textbf{14.5} & 9.8 & 14.0 \tabularnewline
\hline 
Viv1(4Hz) &  26.3 & 17.9 & 12.9 & 17.6 & 13.0 & \textbf{14.4} \tabularnewline
\hline 
Viv2(8Hz) &  28.7 & 8.7 & 6.0 & 34.8 & 9.2 & \textbf{19.0} \tabularnewline
\hline 
Viv2(4Hz) &  27.2 & 10.1 & 7.9 & 36.0 & 9.9 & \textbf{11.2} \tabularnewline
\hline 
Viv3(8Hz) &  16.5 & 9.4 & 6.5 & \textbf{17.4} & 7.6 & 14.5 \tabularnewline
\hline
Viv3(4Hz) & 15.8 & 13.0 & 9.8 & \textbf{19.9} & 10.2 & 18.6 \tabularnewline
\hline 
Viv4(8Hz) &  26.7 & 10.5 & 7.8 & 23.3 & 11.4 & \textbf{19.4} \tabularnewline
\hline 
Viv4(4Hz) &   24.9 & 18.2 & 13.5 & 28.3 & 14.7 & \textbf{17.2} \tabularnewline
\hline 
\end{tabular}
\par\end{centering}
\end{table}

In general, the above experiments show the reliability of our proposed approach, especially when the video content is unpredictable. At the same time, the proposed approach does not increase the number of iterations compared to approaches that offer comparable accuracy. The proposed student-t cost function is a good substitute for the existing cost functions used in GME. It can ensure accuracy and maintain relatively low computational load.

\section{Conclusion}

This letter proposes a Global Motion Estimation approach using a parameterised student-t function and demonstrates the approach in the context of a video of a relatively flat scene. The main difference with existing cost function is that it is an interpolation between a very robust function and a function which responds well to poor initialisation. A binary search is used to adapt a parameter to be well suited to the images in the sequence. Thus, the system can be accurate as well as efficient when processing videos. Future work includes improving the efficiency of the parameter adapation method: rather than try different parameters, it should be  possible to estimate the optimal parameter directly from the image content.

\addtolength{\textheight}{-12cm}

\bibliographystyle{ieeetr}
\bibliography{reference}

\end{document}